\documentclass[twocolumn,showpacs,preprintnumbers,amsmath,amssymb]{revtex4}
\usepackage{graphicx}
\usepackage{dcolumn}
\usepackage{bm}
\begin{document}
\title{Strongly enhanced effective mass in dilute two-dimensional electron systems: System-independent origin}
\author{A.~A. Shashkin, A.~A. Kapustin, E.~V. Deviatov, and V.~T. Dolgopolov}
\affiliation{Institute of Solid State Physics, Chernogolovka, Moscow
District 142432, Russia}
\author{Z.~D. Kvon}
\affiliation{Institute of Semiconductor Physics, Novosibirsk, Russia}
\begin{abstract}
We measure the effective mass in a dilute two-dimensional electron
system in (111)-silicon by analyzing temperature dependence of the
Shubnikov-de~Haas oscillations in the low-temperature limit. A strong
enhancement of the effective mass with decreasing electron density is
observed. The mass renormalization as a function of the interaction
parameter $r_s$ is in good agreement with that reported for
(100)-silicon, which shows that the relative mass enhancement is
system- and disorder-independent being determined by
electron-electron interactions only.
\end{abstract}
\maketitle

The ground state of an ideal, strongly interacting two-dimensional
(2D) electron system is expected to be Wigner crystal
\cite{wigner34}. The interaction strength is characterized by the
Wigner-Seitz radius, $r_s$, which is equal in the single-valley case
to the ratio between the Coulomb energy and the Fermi energy,
$E_C/E_F$. This ratio is proportional to $n_s^{-1/2}$ and therefore
increases with decreasing electron density, $n_s$. According to
numeric simulations \cite{tanatar89}, Wigner crystallization is
expected at $r_s\approx35$. The refined numeric simulations
\cite{attaccalite02} have predicted that prior to the
crystallization, in the range of the interaction parameter
$25\lesssim r_s\lesssim35$, the ground state of the system is a
strongly correlated ferromagnetic Fermi liquid; yet, other
intermediate phases may also exist \cite{spivak03}. At $r_s\sim1$,
the electron liquid is expected to be paramagnetic, with the
effective mass, $m$, and Land\'e $g$ factor renormalized by
interactions. It was not until recently that qualitative deviations
from the weakly-interacting Fermi liquid behavior (in particular, the
drastic increase of the effective electron mass with decreasing
electron density) have been found in strongly correlated 2D electron
systems ($r_s\gtrsim10$) \cite{review,shashkin06,anissimova06}.

The strongest many-body effects have been observed in (100)-silicon
metal-oxide-semiconductor-field-effect-transistors (MOSFETs). Due to
this, there has been recently a revival of interest to (111)-silicon
MOSFETs \cite{estibals04,eng05,eng07}. Although the latter electron
system has been under study for quite a long time, the main
experimental results were obtained some decades ago (see, e.g.,
Ref.~\cite{neugebauer75}), when the knowledge of the 2D electron
systems left to be desired. Electron densities and temperatures used
in experiments were not low enough and the experimental accuracy
achieved for low-mobility samples was not high enough.

In this paper, we report accurate measurements of the effective mass
in a dilute 2D electron system in (111)-silicon by analyzing
temperature dependence of the weak-field Shubnikov-de~Haas (SdH)
oscillations in the low-temperature limit. We find that the effective
mass is strongly increased at low electron densities. The
renormalization of the effective mass as a function of the
interaction parameter $r_s$ agrees well with that found in
(100)-silicon MOSFETs, although the effective masses in bulk silicon
for both orientations are different by a factor of about two. Also,
our (111)-samples have much higher level of disorder than (100)-ones.
This gives evidence that the relative mass enhancement is system- and
disorder-independent and is determined by electron-electron
interactions only.

Measurements were made in an Oxford dilution refrigerator with a base
temperature of $\approx 30$~mK on (111)-silicon MOSFETs similar to
those previously used in Ref.~\cite{estibals04}. Samples had the Hall
bar geometry with width 400~$\mu$m equal to the distance between the
potential probes. Application of a dc voltage to the gate relative to
the contacts allowed one to control the electron density. Oxide
thickness was equal to 1540~\AA . In highest-mobility samples, the
normal of the sample surface was tilted from [111]- toward
[110]-direction by a small angle of $8^\circ$. Anisotropy for
electron transport in such samples does not exceed 5\% at
$n_s=3\times10^{11}$~cm$^{-2}$ and increases weakly with electron
density, staying below 25\% at $n_s=3\times10^{12}$~cm$^{-2}$, as has
been determined in independent experiments. The resistance, $R_{xx}$,
was measured by a standard 4-terminal technique at a low frequency
(5--11~Hz) to minimize the out-of-phase signal. Excitation current
was kept low enough (3--40~nA) to ensure that measurements were taken
in the linear regime of response. SdH oscillations were studied on
two samples, and very similar results were obtained. In particular,
the extracted values of the effective mass were coincident within our
experimental uncertainty. Below, we show results obtained on a sample
with a peak electron mobility close to 2500~cm$^2$/Vs at $T=1.5$~K.

\begin{figure}
\scalebox{0.39}{\includegraphics{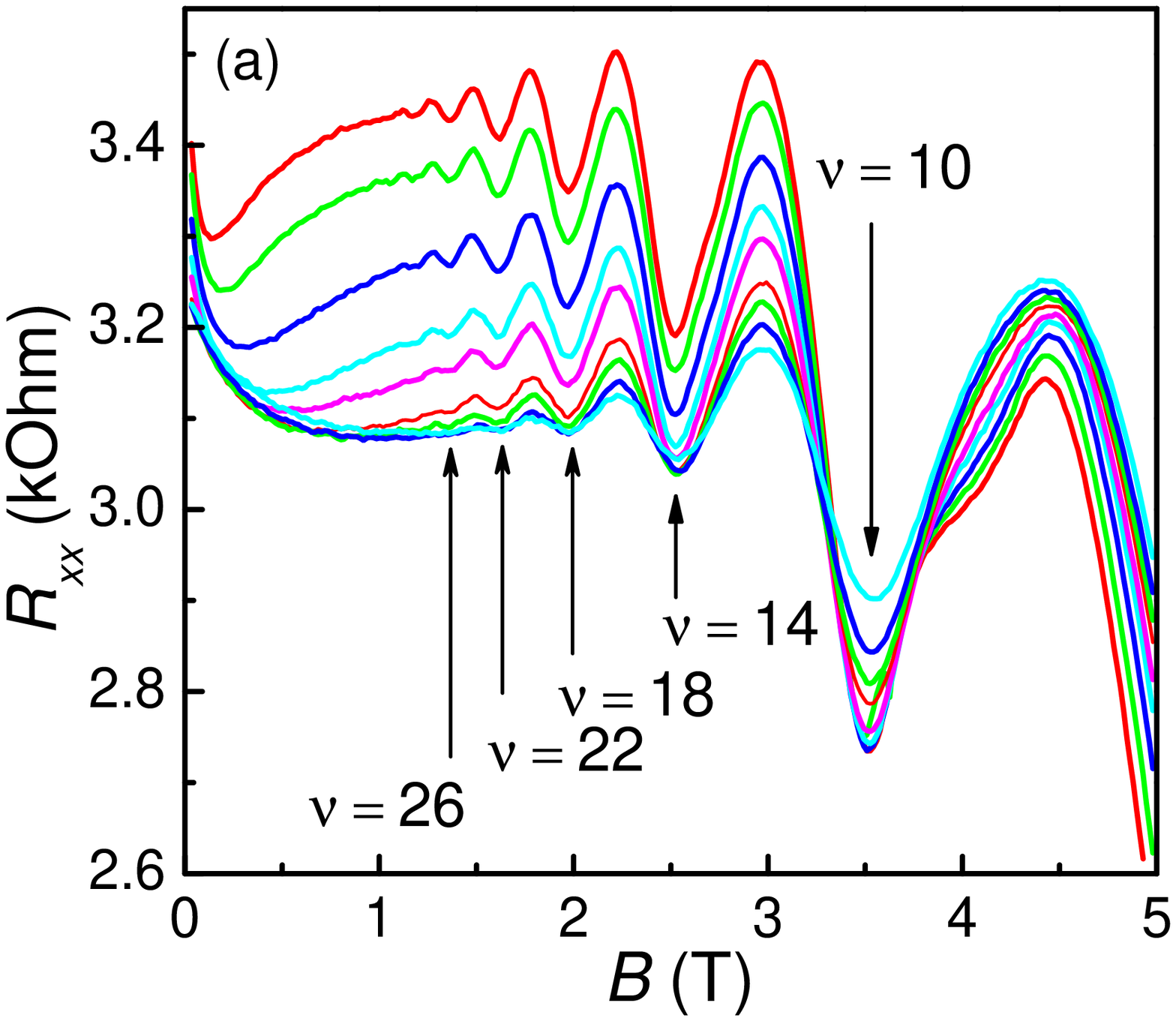}}\vspace{0.14in}
\scalebox{0.39}{\includegraphics{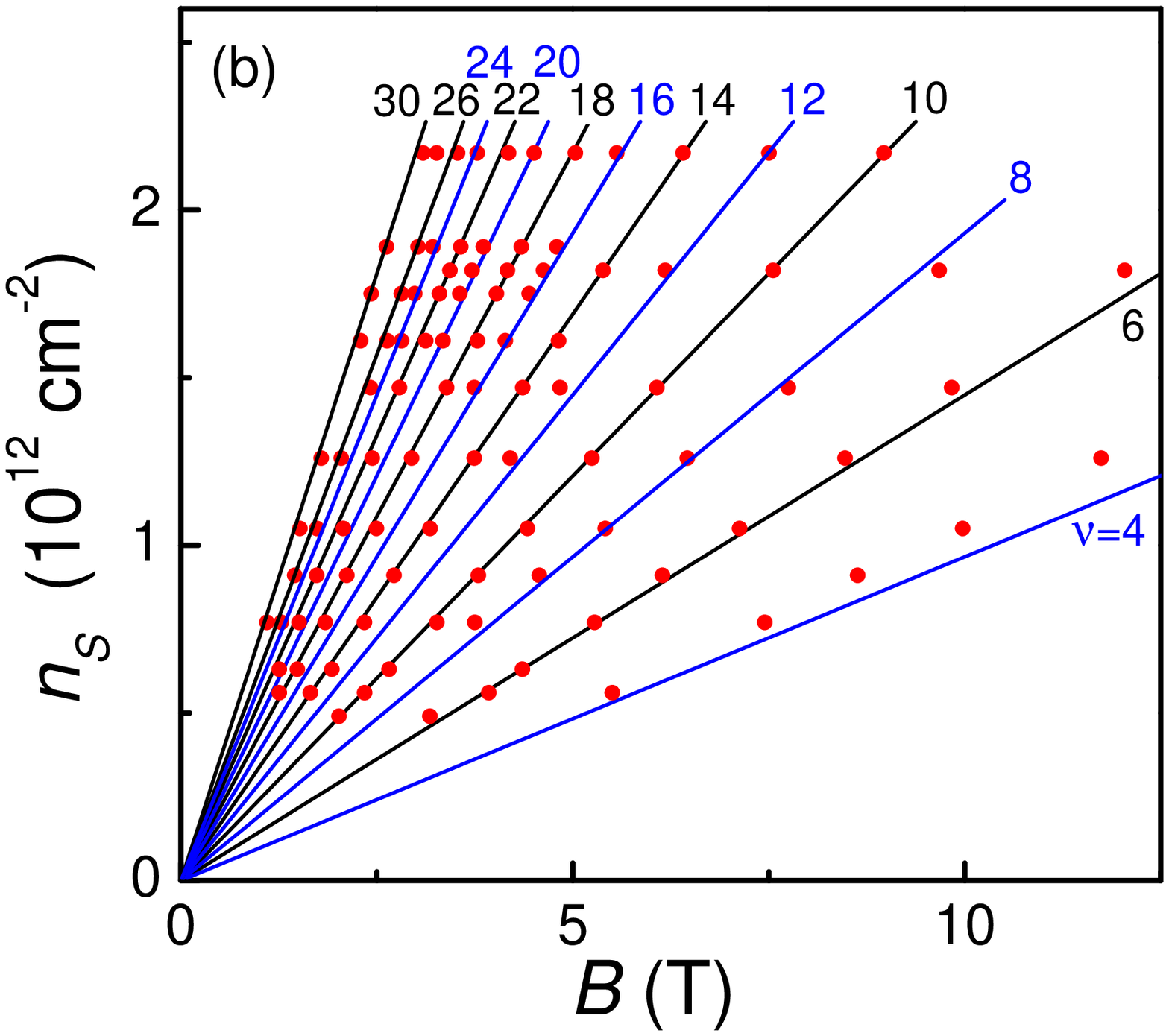}}\vspace{0.14in}
\scalebox{0.39}{\includegraphics{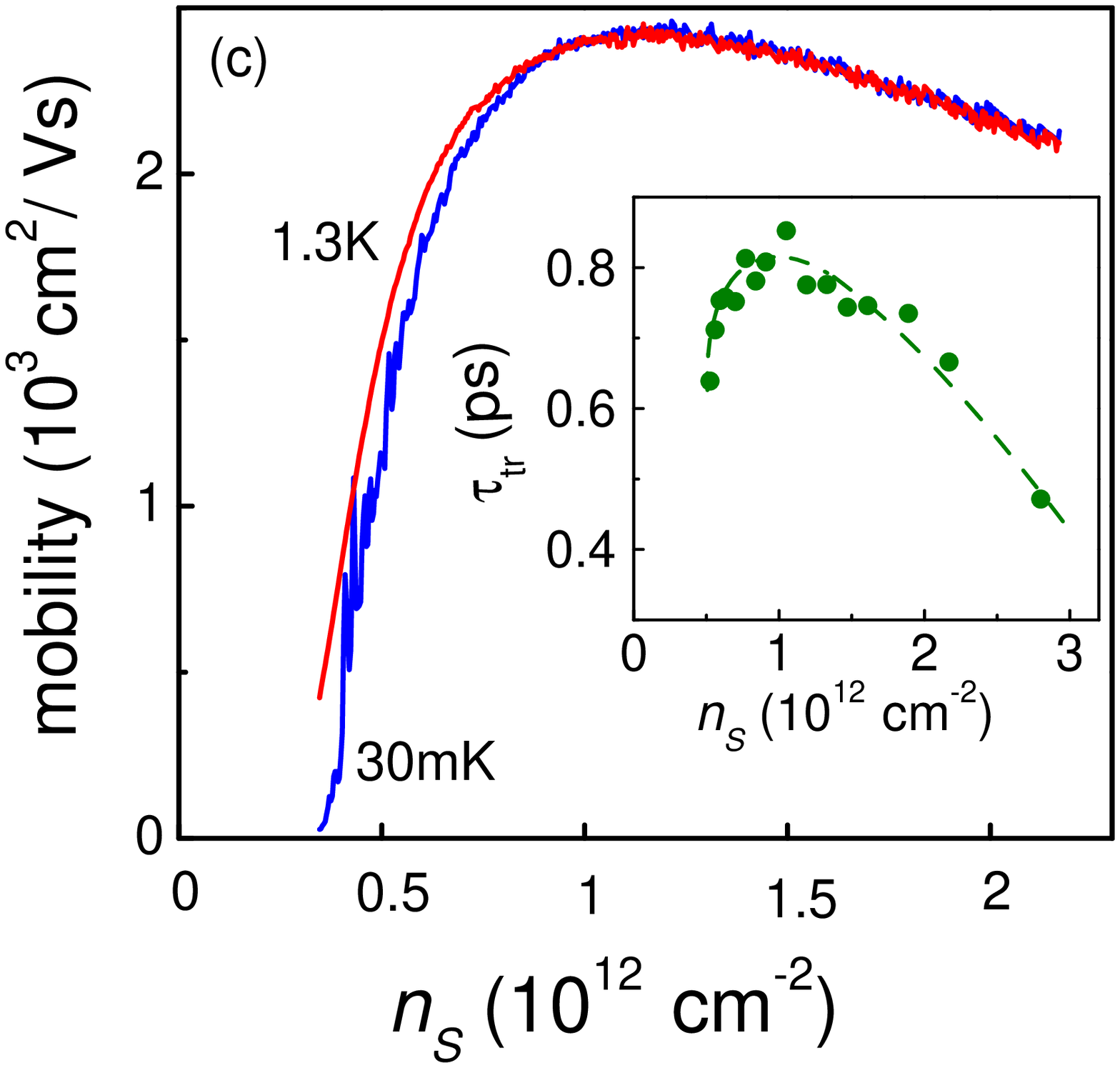}}
\caption{\label{fig1} (a)~Shubnikov-de~Haas oscillations in the
(111)-Si MOSFET at $n_s=8.4\times10^{11}$~cm$^{-2}$ for the following
temperatures (from top to bottom): 0.03, 0.12, 0.2, 0.3, 0.38, 0.47,
0.55, 0.62, and 0.75~K. (b)~Positions of the SdH oscillation minima
in the ($B,n_s$) plane (dots) and the expected positions of the
cyclotron and spin minima calculated according to the formula
$n_s=\nu eB/hc$ (solid lines). (c)~Dependence of the zero-field
mobility on electron density at different temperatures. Inset:
transport scattering time versus $n_s$ evaluated from zero-field
mobility, taking account of the mass renormalization. The dashed line
is a guide to the eye.}
\end{figure}

In Fig.~\ref{fig1}(a), we show the magnetoresistance $R_{xx}(B)$ for
$n_s=8.4\times10^{11}$~cm$^{-2}$ at different temperatures. In weak
magnetic fields, the SdH oscillation period corresponds to a change
of the filling factor $\nu=n_shc/eB$ by $\Delta\nu=4$ \cite{remark},
which indicates that both the spin and valley degeneracies are equal
to $g_s=g_v=2$. The fact that the valley degeneracy is equal to
$g_v=2$, rather than $g_v=6$, is a long-standing problem which lacks
a definite answer so far \cite{estibals04}.

In Fig.~\ref{fig1}(b), we plot positions of the resistance minima in
the ($B,n_s$) plane. The symbols are the experimental data and the
lines are the expected positions of the cyclotron and spin minima
calculated according to the formula $n_s=\nu eB/hc$. The resistance
minima are seen at $\nu=6$, 10, 14, 18, 22, 26, and 30 corresponding
to spin splittings and at $\nu=4$, 8, 12, 16, 20, 24, and 28
corresponding to cyclotron gaps. The valley splitting is not seen at
low electron densities/weak magnetic fields, and the even numbers of
the SdH oscillation minima confirm the valley degeneracy equal to
$g_v=2$. The spin minima extend to appreciably lower electron
densities than the cyclotron minima: behavior that is similar to that
observed in (100)-silicon MOSFETs \cite{kravchenko00}. This reveals
that at the lowest electron densities, the spin splitting is close to
the cyclotron splitting, i.e., the product $gm$ is strongly enhanced
(by a factor of about three).

We would like to emphasize that unlike (100)-silicon MOSFETs with
mobilities in excess of $\approx2\times10^4$~cm$^2$/Vs, the metallic
temperature dependence of the $B=0$ resistance is not observed below
$T=1.3$~K in our samples, as is evident from Fig.~\ref{fig1}(c) which
shows zero-field mobility as a function of electron density at
different temperatures.

\begin{figure}
\scalebox{0.38}{\includegraphics{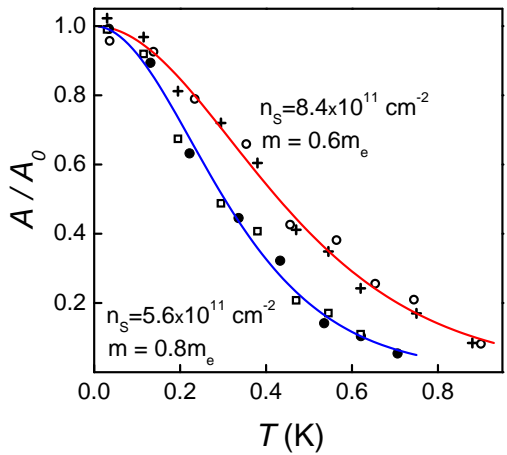}}
\caption{\label{fig2} Change of the amplitude of the weak-field SdH
oscillations with temperature at different electron densities for
magnetic fields $B_1=1.44$~T (dots), $B_2=1.64$~T (squares) and
$B_1=1.45$~T (open circles), $B_2=1.74$~T (crosses). The value of $T$
for the $B_1$ data is multiplied by the ratio $B_2/B_1$. The solid
lines are fits using Eq.~(\ref{A}).}
\end{figure}

A typical temperature dependence of the amplitude, $A$, of the
weak-field (sinusoidal) SdH oscillations for the normalized
resistance, $R_{xx}/R_0$ (where $R_0$ is the average resistance), is
displayed in Fig.~\ref{fig2}. To determine the effective mass, we use
the method of Ref.~\cite{smith72} extending it to low electron
densities and temperatures. We fit the data for $A(T)$ using the
formula
\begin{eqnarray}A(T)&=&A_0\frac{2\pi^2k_BT/\hbar\omega_c}{\sinh(2\pi^2k_BT/\hbar\omega_c)},
\nonumber\\A_0&=&4\exp(-2\pi^2k_BT_D/\hbar\omega_c),\label{A}\end{eqnarray}
where $\omega_c=eB/mc$ is the cyclotron frequency and $T_D$ is the
Dingle temperature \cite{lifshitz55,shashkin03}. The latter is
related to the level width through the expression $T_D=\hbar/2\pi
k_B\tau$, where $\tau$ is the quantum scattering time \cite{ando82}.
In principle, temperature-dependent $\tau$ may influence damping of
the SdH oscillations with temperature. In our experiment, however,
possible corrections to the mass value caused by the temperature
dependence of $\tau$ (and hence $T_D$) are within the experimental
uncertainty which is estimated by data dispersion at about 10\%. Note
that the amplitude of the SdH oscillations follows the calculated
curve down to the lowest achieved temperatures, which confirms that
the electrons were in a good thermal contact with the bath and were
not overheated. Applicability of Eq.~(\ref{A}) to strongly
interacting 2D electron systems is justified by the coincidence of
the enhanced mass values obtained in (100)-silicon MOSFETs using a
number of independent measurement methods including this one
\cite{anissimova06,shashkin03,shashkin02}.

\begin{figure}[b]\vspace{-0.16in}
\scalebox{0.38}{\includegraphics{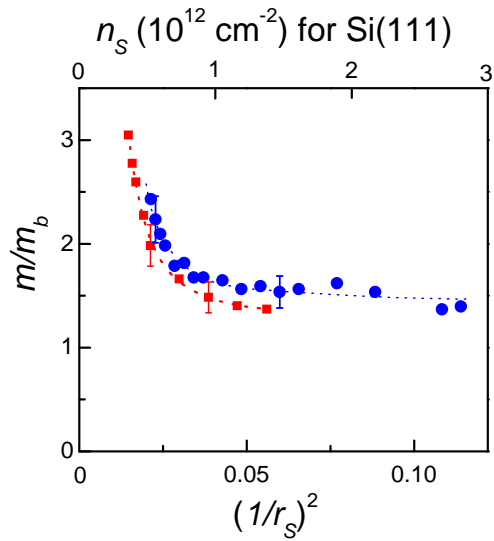}}
\caption{\label{fig3} The effective mass (dots) in units of $m_b$ as
a function of $(1/r_s)^2\propto n_s$. Also shown by squares is the
data obtained in (100)-Si MOSFETs \cite{shashkin03}. The dashed lines
are guides to the eye.}
\end{figure}

In Fig.~\ref{fig3}, we show the so-determined effective mass in units
of the cyclotron mass in bulk silicon, $m_b=0.358m_e$ (where $m_e$ is
the free electron mass), as a function of $(1/r_s)^2\propto n_s$.
(For the two-valley case the ratio $E_C/E_F$ that determines the
system behavior is twice as large as the Wigner-Seitz radius $r_s$.
To avoid confusion, below we will still use the Wigner-Seitz radius
with the average dielectric constant of 7.7 as the interaction
parameter.) The effective mass sharply increases with decreasing
electron density, its enhancement at low $n_s$ being consistent with
that of $gm$. The mass renormalization $m/m_b$ versus the interaction
parameter $r_s$ is coincident within the experimental uncertainty
with that found in (100)-silicon MOSFETs where $m_b=0.19m_e$ is
approximately twice as small and the peak mobility is approximately
one order of magnitude as large. Thus, we arrive at a conclusion that
the relative mass enhancement is determined by $r_s$, being
independent of a 2D electron system. Note that the highest accessible
$r_s$ is different in different 2D electron systems.

\begin{figure}
\scalebox{0.38}{\includegraphics{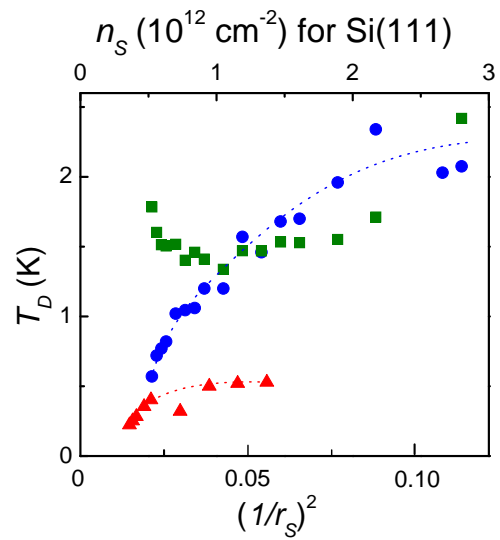}}
\caption{\label{fig4} Behavior of the Dingle temperature extracted
from SdH oscillations (dots) and that calculated from the transport
scattering time (squares) with $(1/r_s)^2\propto n_s$. Also shown by
triangles is the data obtained from SdH oscillations in (100)-Si
MOSFETs \cite{shashkin03}. The dashed lines are guides to the eye.}
\end{figure}

In Fig.~\ref{fig4}, we compare the extracted Dingle temperature $T_D$
with that recalculated from the electron lifetime, $\tau_{tr}$, that
is evaluated from zero-field mobility, taking into account the mass
renormalization \cite{shashkin02} (inset to Fig.~\ref{fig1}(c)).
Although the quantum scattering time $\tau$ defining the Dingle
temperature is in general different from the transport scattering
time $\tau_{tr}$, the two $T_D(n_s)$ dependences are consistent with
each other at $(1/r_s)^2>0.04$, indicating dominant large-angle
scattering \cite{ando82}. The measured value of $T_D$ at the same
$r_s$ is considerably larger in (111)-silicon than in (100)-silicon
MOSFETs and, therefore, the Dingle temperature increases with
disorder. In both electron systems, the Dingle temperature decreases
with decreasing electron density. It is worth noting that the
different behavior of the measured and recalculated $T_D(n_s)$
dependences at low electron densities (at $(1/r_s)^2<0.04$) is
consistent with predictions of the scattering theory of
Ref.~\cite{gold88}. There, it was shown that multiple scattering
effects lead to a decrease of the ratio $\tau_{tr}/\tau$ at low
electron densities so that this ratio tends to zero at the
metal-insulator transition.

We now discuss the results obtained for the effective mass. We stress
that the strongly increased mass is observed in a dilute 2D electron
system with relatively high disorder, as inferred from both the
relatively low zero-field mobility and the absence of metallic
temperature dependence of zero-field resistance. Moreover, the
disorder does not at all influence the relative enhancement of the
mass as a function of the interaction parameter. This allows us to
claim that the mass enhancement is solely caused by electron-electron
interactions. Our results also add confidence that the dilute system
behavior in the regime of the strongly enhanced spin susceptibility
$\chi\propto gm$ --- close to the onset of spontaneous spin
polarization and Wigner crystallization --- is governed by the
effective mass.

The finding that in dilute 2D electron systems the effective mass is
strongly enhanced remains basically unexplained, although there has
been a good deal of theoretical work on the subject (see
Refs.~\cite{review,kravchenko06} and references therein). The latest
theoretical developments include the following. Using a
renormalization group analysis for multi-valley 2D systems, it has
been found that the effective mass dramatically increases at
disorder-dependent density for the metal-insulator transition while
the $g$ factor remains nearly intact \cite{punnoose05}. However, the
prediction of disorder-dependent effective mass is not confirmed by
our data. In the Fermi-liquid-based model of Ref.~\cite{khodel05}, a
flattening at the Fermi energy in the spectrum that leads to a
diverging effective mass has been predicted. Still, the expected
dependence of the effective mass on temperature is not consistent
with experimental findings.

Finally, we would like to note that moderate enhancements of the
effective mass $m\approx1.5m_b$ have been determined in 2D electron
systems of AlAs quantum wells and GaAs/AlGaAs heterostructures
because the lowest accessible densities are still too high
\cite{vakili04,tan05,remark1}. While the theories of, e.g.,
Refs.~\cite{zhang05,asgari06} are capable of describing the
experimental $m(n_s)$ dependence at $r_s\sim1$, their validity at
larger values of the interaction parameter is a problem.

In summary, we have found that in a dilute 2D electron system in
(111)-silicon, the effective mass sharply increases at low electron
densities. The mass renormalization versus the interaction parameter
$r_s$ is in good agreement with that reported for (100)-silicon
MOSFETs. This gives evidence that the relative mass enhancement is
system- and disorder-independent and is solely determined by
electron-electron interactions. The obtained results show that the
dilute system behavior in the regime of the strongly enhanced spin
susceptibility is governed by the effective mass. The particular
mechanism underlying the effective mass enhancement remains to be
seen.

We gratefully acknowledge discussions with E. Abrahams, A. Gold,
S.~V. Kravchenko, and A. Punnoose. This work was supported by the
RFBR, RAS, and the Programme ``The State Support of Leading
Scientific Schools''.


\begin{thebibliography}{apssamp}
\bibitem{remark} On higher-mobility Si(111) samples of
Ref.~\cite{eng07}, a periodicity $\Delta\nu=8$ was observed in weak
magnetic fields.
\bibitem{remark1} The variation of the extracted $m$ in different
samples and different cooldowns, found in Ref.~\cite{vakili04}, was
attributed to sample inhomogeneities, as manifested by the spikes on
the dependence of the electron mobility on $n_s$.
\end{thebibliography}
\end{document}